\documentclass[conference]{IEEEtran}

\usepackage{epsfig,amsmath,amssymb,epsf,amsthm,scalefnt,multirow,subcaption}
\usepackage[labelsep=period]{caption}
\usepackage{xcolor}
\usepackage{float}
\usepackage{cite}
\usepackage{psfrag}
\usepackage{algorithm}
\usepackage{algpseudocode}
\usepackage[top=0.75in, bottom=1in, left=0.65in, right=0.65in]{geometry}

\algrenewcommand\algorithmicrequire{\textbf{Input:}}
\algrenewcommand\algorithmicensure{\textbf{Output:}}

\newtheorem*{lemma*}{Lemma}
\newtheorem{observation}{Observation}

\newtheorem{conjecture}{Conjecture}

\begin{document}

\title{\huge{Gradient Pursuit-Based Channel Estimation for MmWave Massive MIMO Systems with One-Bit ADCs}}

\author{\IEEEauthorblockN{In-soo Kim and Junil Choi}
\IEEEauthorblockA{Department of Electrical Engineering\\
POSTECH\\
Pohang, Korea\\
Email: \{insookim, junil\}@postech.ac.kr}}

\maketitle

\begin{abstract}
In this paper, channel estimation for millimeter wave (mmWave) massive multiple-input multiple-output (MIMO) systems with one-bit analog-to-digital converters (ADCs) is considered. In the mmWave band, the number of propagation paths is small, which results in sparse virtual channels. To estimate sparse virtual channels based on the maximum a posteriori (MAP) criterion, sparsity-constrained optimization comes into play. In general, optimizing objective functions with sparsity constraints is NP-hard because of their combinatorial complexity. Furthermore, the coarse quantization of one-bit ADCs makes channel estimation a challenging task. In the field of compressed sensing (CS), the gradient support pursuit (GraSP) and gradient hard thresholding pursuit (GraHTP) algorithms were proposed to approximately solve sparsity-constrained optimization problems iteratively by pursuing the gradient of the objective function via hard thresholding. The accuracy guarantee of these algorithms, however, breaks down when the objective function is ill-conditioned, which frequently occurs in the mmWave band. To prevent the breakdown of gradient pursuit-based algorithms, the band maximum selecting (BMS) technique, which is a hard thresholder selecting only the ``band maxima,'' is applied to GraSP and GraHTP to propose the BMSGraSP and BMSGraHTP algorithms in this paper.
\end{abstract}

\section{Introduction}
The wide bandwidth of the millimeter wave (mmWave) band provides high data rates, resulting in a significant performance gain \cite{6894453, 6515173, 6736746, 6732923}. Also, the small wavelength enables the use of large arrays at the transmitter and receiver, which is widely known as massive multiple-input multiple-output (MIMO) systems. The power consumption of an analog-to-digital converter (ADC), however, increases linearly with the sampling rate and exponentially with the ADC resolution, which makes high-resolution ADCs impractical for mmWave massive MIMO systems \cite{1550190}. To reduce the impractically high power consumption, one possible solution is to use low-resolution ADCs, which recently gained popularity \cite{7600443, 7307134, 7420605, 7894211}. In this paper, we consider the extreme scenario of mmWave massive MIMO systems with one-bit ADCs.

The number of propagation paths in the mmWave band is small, which naturally leads to sparse virtual channels. Therefore, channel estimation for mmWave massive MIMO systems with one-bit ADCs can be formulated as sparse signal recovery with quantized measurements. In \cite{8310593, 8320852, 8171203}, compressed sensing-based (CS-based) signal recovery techniques for mmWave massive MIMO systems with low-resolution ADCs were proposed. In \cite{8310593}, a sparse Bayesian learning-based (SBL-based) approximate maximum a posteriori (MAP) channel estimator was proposed, which sought the channel estimate using the variational Bayesian (VB) method. The generalized expectation consistent signal recovery (GEC-SR) \cite{8320852} and generalized approximate message passing (GAMP) \cite{8171203} algorithms are iterative approximate minimum mean squared error (MMSE) estimators, which are based on the turbo principle and loopy belief propagation (BP). However, the accuracy guarantee of these algorithms breaks down when the sensing matrix is ill-conditioned, which occurs when the angular domain grid resolution of the virtual channel representation is too high.

In this paper, we propose gradient pursuit-based iterative approximate MAP channel estimators for mmWave massive MIMO systems with one-bit ADCs. In the mmWave band, the MAP channel estimation framework can be cast into a sparsity-constrained optimization problem. To approximately solve such problem iteratively, we adopt the gradient support pursuit (GraSP) \cite{bahmani2013greedy} and gradient hard thresholding pursuit (GraHTP) \cite{yuan2017gradient} algorithms, which are gradient pursuit-based CS techniques. Similar to the aforementioned CS-based algorithms, however, these algorithms break down when the grid resolution is too high because the resulting objective function is ill-conditioned. To prevent such breakdown, we propose the band maximum selecting (BMS) technique, which is a hard thresholder selecting only the ``band maxima.'' The BMS technique is then applied to GraSP and GraHTP, which results in the proposed BMSGraSP and BMSGraHTP algorithms. According to the simulation results, BMSGraSP and BMSGraHTP outperform other channel estimators including GAMP, which shows the superiority of our proposed techniques.

\textbf{Notation:} $a$, $\mathbf{a}$, and $\mathbf{A}$ denote a scalar, vector, and matrix. The transpose, conjugate transpose, and conjugate of $\mathbf{A}$ are denoted as $\mathbf{A}^{\mathrm{T}}$, $\mathbf{A}^{\mathrm{H}}$, and $\overline{\mathbf{A}}$. The Kronecker product of $\mathbf{A}$ and $\mathbf{B}$ is denoted as $\mathbf{A}\otimes\mathbf{B}$. The support of $\mathbf{a}$ is denoted as $\mathrm{supp}(\mathbf{a})$, which is formed by collecting all of the indices of the nonzero elements of $\mathbf{a}$. The best $s$-term approximation of $\mathbf{a}$ is denoted as $\mathbf{a}|_{s}$, which is obtained by leaving only the $s$ largest elements of $\mathbf{a}$ and hard thresholding other elements to $0$. For a set $\mathcal{A}$, the vector obtained by leaving only the elements of $\mathbf{a}$ indexed by $\mathcal{A}$ and hard thresholding the remaining elements to $0$ is denoted as $\mathbf{a}|_{\mathcal{A}}$. The absolute value of a scalar $a$ and cardinality of a set $\mathcal{A}$ are denoted as $|a|$ and $|\mathcal{A}|$. The standard normal PDF and CDF are denoted as $\phi(x)=\frac{1}{\sqrt{2\pi}}e^{-\frac{x^{2}}{2}}$ and $\Phi(x)=\int_{-\infty}^{x}\frac{1}{\sqrt{2\pi}}e^{-\frac{y^{2}}{2}}dy$. The inverse Mills ratio function is defined as $\lambda(x)=\frac{\phi(x)}{\Phi(x)}$. The element-wise matrix multiplication and division are denoted as $\odot$ and $\oslash$. The element-wise standard normal PDF, CDF, and inverse Mills ratio function are $\phi(\mathbf{x})$, $\Phi(\mathbf{x})$, and $\lambda(\mathbf{x})=\phi(\mathbf{x})\oslash\Phi(\mathbf{x})$.

\section{System Model}\label{system_model}
Consider a mmWave massive MIMO system with one-bit ADCs, whose $N$-antenna transmitter and $M$-antenna receiver are equipped with uniform linear arrays (ULAs). In the channel estimation phase, a training sequence of length $T$ is transmitted to the receiver. By collecting the signals over the $T$ time slots, the received signal $\mathbf{Y}\in\mathbb{C}^{M\times T}$ is
\begin{equation}
\mathbf{Y}=\sqrt{\rho}\mathbf{H}\mathbf{S}+\mathbf{N}
\end{equation}
where $\mathbf{H}\in\mathbb{C}^{M\times N}$ is the channel, $\mathbf{S}\in\mathbb{C}^{N\times T}$ is the training sequence, whose columns obey the $2$-norm constraint of $\sqrt{N}$, and $\mathbf{N}\in\mathbb{C}^{M\times T}$ is the additive white Gaussian noise (AWGN), which is distributed as $\mathrm{vec}(\mathbf{N})\sim\mathcal{CN}(\mathbf{0}_{MT}, \mathbf{I}_{MT})$. The signal-to-noise ratio (SNR) is defined as $\rho$. In the mmWave band, $\mathbf{H}$ contains a small number of paths where each path is associated with its path gain, angle-of-arrival (AoA), and angle-of-departure (AoD) \cite{6965800}. Then, $\mathbf{H}$ is
\begin{equation}
\mathbf{H}=\sum_{\ell=1}^{L}\alpha_{\ell}\mathbf{a}_{\mathrm{RX}}(\theta_{\mathrm{RX}, \ell})\mathbf{a}_{\mathrm{TX}}(\theta_{\mathrm{TX}, \ell})^{\mathrm{H}}
\end{equation}
where $L$ is the number of paths, $\alpha_{\ell}\sim\mathcal{CN}(0, 1)$ is the $\ell$-th path gain, and $\theta_{\mathrm{RX}, \ell}\sim\mathrm{unif}([-\pi/2, \pi/2])$ and $\theta_{\mathrm{TX}, \ell}\sim\mathrm{unif}([-\pi/2, \pi/2])$ are the $\ell$-th AoA and AoD, which are independent. The steering vectors $\mathbf{a}_{\mathrm{RX}}(\theta_{\mathrm{RX}, \ell})\in\mathbb{C}^{M}$ and $\mathbf{a}_{\mathrm{TX}}(\theta_{\mathrm{TX}, \ell})\in\mathbb{C}^{N}$ are
\begin{align}
\mathbf{a}_{\mathrm{RX}}(\theta_{\mathrm{RX}, \ell})&=\frac{1}{\sqrt{M}}\begin{bmatrix}1&\cdots&e^{-j\pi(M-1)\sin(\theta_{\mathrm{RX}, \ell})}\end{bmatrix}^{\mathrm{T}},\\
\mathbf{a}_{\mathrm{TX}}(\theta_{\mathrm{TX}, \ell})&=\frac{1}{\sqrt{N}}\begin{bmatrix}1&\cdots&e^{-j\pi(N-1)\sin(\theta_{\mathrm{TX}, \ell})}\end{bmatrix}^{\mathrm{T}},
\end{align}
whose inter-element spacings are half-wavelength. The one-bit quantized received signal $\hat{\mathbf{Y}}\in\mathbb{C}^{M\times T}$ is
\begin{align}\label{quantized_received_signal}
\hat{\mathbf{Y}}&=\mathrm{Q}(\mathbf{Y})\notag\\
                &=\mathrm{Q}(\sqrt{\rho}\mathbf{H}\mathbf{S}+\mathbf{N})
\end{align}
where $\mathrm{Q}(\cdot)$ is the zero threshold one-bit quantization function defined as
\begin{equation}\label{quantization}
\mathrm{Q}(\mathbf{Y})=\mathrm{sign}(\mathrm{Re}(\mathbf{Y}))+j\mathrm{sign}(\mathrm{Im}(\mathbf{Y}))
\end{equation}
with $\mathrm{sign}(\cdot)$ the element-wise sign function.

The virtual channel representation \cite{1033686} of $\mathbf{H}$ is
\begin{equation}\label{virtual_channel_representation}
\mathbf{H}=\mathbf{A}_{\mathrm{RX}}\mathbf{X}^{*}\mathbf{A}_{\mathrm{TX}}^{\mathrm{H}}
\end{equation}
where $\mathbf{A}_{\mathrm{RX}}\in\mathbb{C}^{M\times B_{\mathrm{RX}}}$ and $\mathbf{A}_{\mathrm{TX}}\in\mathbb{C}^{N\times B_{\mathrm{TX}}}$ are overcomplete discrete Fourier transform (DFT) matrices, and $\mathbf{X}^{*}\in\mathbb{C}^{B_{\mathrm{RX}}\times B_{\mathrm{TX}}}$ is the virtual channel with $B_{\mathrm{RX}}\geq M$ and $B_{\mathrm{TX}}\geq N$. For notational simplicity, we define $B=B_{\mathrm{RX}}B_{\mathrm{TX}}$. To facilitate the analysis, \eqref{quantized_received_signal} is vectorized after plugging in \eqref{virtual_channel_representation} as
\begin{align}
\hat{\mathbf{y}}&=\mathrm{Q}(\mathbf{y})\notag\\
                &=\mathrm{Q}(\sqrt{\rho}\mathbf{A}\mathbf{x}^{*}+\mathbf{n})
\end{align}
where $\hat{\mathbf{y}}=\mathrm{vec}(\hat{\mathbf{Y}})$, $\mathbf{y}=\mathrm{vec}(\mathbf{Y})$, $\mathbf{A}=\mathbf{S}^{\mathrm{T}}\overline{\mathbf{A}}_{\mathrm{TX}}\otimes\mathbf{A}_{\mathrm{RX}}=\begin{bmatrix}\mathbf{a}_{1}&\cdots&\mathbf{a}_{B}\end{bmatrix}$, $\mathbf{x}^{*}=\mathrm{vec}(\mathbf{X}^{*})=\begin{bmatrix}x_{1}^{*}&\cdots&x_{B}^{*}\end{bmatrix}^{\mathrm{T}}$, and $\mathbf{n}=\mathrm{vec}(\mathbf{N})$.

To build the MAP channel estimation framework, we start with formulating the likelihood function based on the real forms of $\hat{\mathbf{y}}$, $\mathbf{A}$, and $\mathbf{x}^{*}$, which are
\begin{align}
\hat{\mathbf{y}}_{\mathrm{R}}&=\begin{bmatrix}\mathrm{Re}(\hat{\mathbf{y}})^{\mathrm{T}}&\mathrm{Im}(\hat{\mathbf{y}})^{\mathrm{T}}\end{bmatrix}^{\mathrm{T}}\notag\\
                             &=\begin{bmatrix}\hat{y}_{\mathrm{R}, 1}&\cdots&\hat{y}_{\mathrm{R}, 2MT}\end{bmatrix}^{\mathrm{T}},\\
      \mathbf{A}_{\mathrm{R}}&=\begin{bmatrix}\mathrm{Re}(\mathbf{A})&-\mathrm{Im}(\mathbf{A})\\\mathrm{Im}(\mathbf{A})&\mathrm{Re}(\mathbf{A})\end{bmatrix}\notag\\
                             &=\begin{bmatrix}\mathbf{a}_{\mathrm{R}, 1}&\cdots&\mathbf{a}_{\mathrm{R}, 2MT}\end{bmatrix}^{\mathrm{T}},\\
  \mathbf{x}_{\mathrm{R}}^{*}&=\begin{bmatrix}\mathrm{Re}(\mathbf{x}^{*})^{\mathrm{T}}&\mathrm{Im}(\mathbf{x}^{*})^{\mathrm{T}}\end{bmatrix}^{\mathrm{T}}
\end{align}
where the complex forms and their real forms are used interchangeably in the sequel. For example, $\mathbf{x}^{*}$ and $\mathbf{x}_{\mathrm{R}}^{*}$ represent the same entity. Then, the log-likelihood function $f(\mathbf{x})$ is \cite{7439790}
\begin{align}
f(\mathbf{x})&=\log\mathrm{Pr}\begin{bmatrix}\hat{\mathbf{y}}=\mathrm{Q}(\sqrt{\rho}\mathbf{A}\mathbf{x}+\mathbf{n})\mid\mathbf{x}\end{bmatrix}\notag\\
             &=\sum_{i=1}^{2MT}\log\Phi(\sqrt{2\rho}\hat{y}_{\mathrm{R}, i}\mathbf{a}_{\mathrm{R}, i}^{\mathrm{T}}\mathbf{x}_{\mathrm{R}}).
\end{align}
However, $\mathbf{x}^{*}$ is approximately sparse due to the leakage effect \cite{7094595}, which complicates the channel estimation problem. To formulate the MAP channel estimation framework, we assume that $\mathbf{x}^{*}$ is $L$-sparse with independent and identically distributed (i.i.d.) $\mathcal{CN}(0, 1)$ elements, whose locations are uniformly distributed in $\{1, \dots, B\}$. The mismatch between our assumption on $\mathbf{x}^{*}$ and its true distribution reduces as we increase $B_{\mathrm{RX}}$ and $B_{\mathrm{TX}}$ due to the reduced leakage effect. In Section \ref{simulation_results}, it is shown that sufficiently large $B_{\mathrm{RX}}$ and $B_{\mathrm{TX}}$ lead to accurate channel estimation because our assumption closely approximates the true distribution of $\mathbf{x}^{*}$. With such assumption on the distribution of $\mathbf{x}^{*}$, the MAP estimate of $\mathbf{x}^{*}$ is
\begin{equation}\label{map}
\underset{\mathbf{x}\in\mathbb{C}^{B}}{\mathrm{argmax}}\ (f(\mathbf{x})+g(\mathbf{x}))\enspace\text{s.t.}\enspace\|\mathbf{x}\|_{0}\leq L
\end{equation}
where $g(\mathbf{x})=-\|\mathbf{x}_{\mathrm{R}}\|^{2}$ is the logarithm of the PDF of $\mathcal{CN}(\mathbf{0}_{B}, \mathbf{I}_{B})$ without the constant factor. In the sequel, the objective function and its gradient in \eqref{map} are denoted as
\begin{align}
       h(\mathbf{x})&=f(\mathbf{x})+g(\mathbf{x}),\\
\nabla h(\mathbf{x})&=\nabla f(\mathbf{x})+\nabla g(\mathbf{x})\notag\\
                    &=\begin{bmatrix}\nabla h(x_{1})&\cdots&\nabla h(x_{B})\end{bmatrix}^{\mathrm{T}},
\end{align}
whose differentiation is with respect to $\mathbf{x}$.

\section{Proposed BMSGraSP and BMSGraHTP algorithms}
In general, solving \eqref{map} is NP-hard due to its sparsity constraint. To approximately optimize sparsity-constrained objective functions iteratively by pursuing the gradient of the objective function, GraSP \cite{bahmani2013greedy} and GraHTP \cite{yuan2017gradient} algorithms were proposed in the field of CS, which generalize the well-known compressive sampling matching pursuit (CoSaMP) \cite{needell2009cosamp} and hard thresholding pursuit (HTP) \cite{foucart2011hard} algorithms to objective functions with arbitrary forms.

To solve \eqref{map}, GraSP and GraHTP update the $L$-sparse current estimate $\hat{\mathbf{x}}$ of $\mathbf{x}^{*}$ roughly as follows at each iteration. The best $L$-term approximation of $\nabla h(\hat{\mathbf{x}})$ is computed via hard thresholding as
\begin{equation}
\nabla h(\hat{\mathbf{x}})|_{L},
\end{equation}
whose support
\begin{equation}\label{support_identification}
\mathcal{I}=\mathrm{supp}(\nabla h(\hat{\mathbf{x}})|_{L})
\end{equation}
is selected as the updated estimate of $\mathrm{supp}(\mathbf{x}^{*})$. This step can be interpreted as support identification, or joint AoA and AoD estimation. Then, the elements corresponding to the identified support are updated by selecting the updated $\hat{\mathbf{x}}$ as
\begin{equation}\label{path_gain_estimation}
\underset{\mathbf{x}\in\mathbb{C}^{B}}{\mathrm{argmax}}\ h(\mathbf{x})\enspace\text{s.t.}\enspace\mathrm{supp}(\mathbf{x})\subseteq\mathcal{I},
\end{equation}
which can be interpreted as path gain estimation. Note that \eqref{path_gain_estimation} is a convex optimization problem because $h(\mathbf{x})$ is concave with its support constraint being convex. The concavity of $h(\mathbf{x})$ follows from the fact that $f(\mathbf{x})$ is concave because $\Phi(\cdot)$ is log-concave; similarly, $g(\mathbf{x})$ is concave due to the convexity of $2$-norm \cite{boyd2004convex}.

The accuracy guarantee of GraSP and GraHTP, however, breaks down when $h(\mathbf{x})$ does not have a stable restricted Hessian \cite{bahmani2013greedy} or is not strongly convex and smooth \cite{yuan2017gradient}. The problem is that for large $B_{\mathrm{RX}}$ and $B_{\mathrm{TX}}$, the resulting $\mathbf{A}$ is ill-conditioned because its columns become highly coherent, which leads to unfavorable $h(\mathbf{x})$. Specifically, with highly coherent $\mathbf{A}$, GraSP and GraHTP are likely to fail to accurately identify the support of $\mathbf{x}^{*}$ from \eqref{support_identification}.

To illustrate how support identification fails from \eqref{support_identification} when $\mathbf{A}$ is highly coherent, first, consider the real form $\nabla h(\mathbf{x_{\mathrm{R}}})$ of $\nabla h(\mathbf{x})$ defined as
\begin{align}\label{gradient}
 &\nabla h(\mathbf{x}_{\mathrm{R}})\notag\\
=&\begin{bmatrix}\mathrm{Re}(\nabla h(\mathbf{x}))^{\mathrm{T}}&\mathrm{Im}(\nabla h(\mathbf{x}))^{\mathrm{T}}\end{bmatrix}^{\mathrm{T}}\notag\\
=&\sum_{i=1}^{2MT}\lambda(\sqrt{2\rho}\hat{y}_{\mathrm{R}, i}\mathbf{a}_{\mathrm{R}, i}^{\mathrm{T}}\mathbf{x}_{\mathrm{R}})\sqrt{2\rho}\hat{y}_{\mathrm{R}, i}\mathbf{a}_{\mathrm{R}, i}-2\mathbf{x}_{\mathrm{R}}\notag\\
=&\mathbf{A}_{\mathrm{R}}^{\mathrm{T}}(\lambda(\sqrt{2\rho}\hat{\mathbf{y}}_{\mathrm{R}}\odot\mathbf{A}_{\mathrm{R}}\mathbf{x}_{\mathrm{R}})\odot\sqrt{2\rho}\hat{\mathbf{y}}_{\mathrm{R}})-2\mathbf{x}_{\mathrm{R}},
\end{align}
which is obtained from $\nabla\log\Phi(\mathbf{a}_{\mathrm{R}}^{\mathrm{T}}\mathbf{x}_{\mathrm{R}})=\lambda(\mathbf{a}_{\mathrm{R}}^{\mathrm{T}}\mathbf{x}_{\mathrm{R}})\mathbf{a}_{\mathrm{R}}$ and $\nabla\|\mathbf{x}_{\mathrm{R}}\|^{2}=2\mathbf{x}_{\mathrm{R}}$. Then, we establish the following observation, which can be checked from directly computing $\nabla h(x_{i})$, whose real and imaginary parts correspond to the $i$-th and $(i+B)$-th elements of $\nabla h(\mathbf{x}_{\mathrm{R}})$.
\begin{observation}\label{observation}
$\nabla h(x_{i})=\nabla h(x_{j})$ if $\mathbf{a}_{i}=\mathbf{a}_{j}$ and $x_{i}=x_{j}$.
\end{observation}
However, Observation \ref{observation} is trivial since $\mathbf{a}_{i}\neq\mathbf{a}_{j}$ unless $i=j$. To introduce a nontrivial observation, we adopt the notion of coherence between $\mathbf{a}_{i}$ and $\mathbf{a}_{j}$, i.e.,
\begin{equation}
\mu(i, j)=\frac{|\mathbf{a}_{i}^{\mathrm{H}}\mathbf{a}_{j}|}{\|\mathbf{a}_{i}\|\|\mathbf{a}_{j}\|},
\end{equation}
which measures the proximity between $\mathbf{a}_{i}$ and $\mathbf{a}_{j}$ \cite{fannjiang2012coherence, 1577905, shrivastava2014asymmetric}. Then, with the $\eta$-coherence band of $i$, which is defined as \cite{fannjiang2012coherence}
\begin{equation}
B_{\eta}(i)=\{j\mid\mu(i, j)\geq\eta\}
\end{equation}
with $\eta\in(0, 1)$, the following conjecture is established for sufficiently large $\eta$.
\begin{conjecture}\label{conjecture}
$\nabla h(x_{i})\approx\nabla h(x_{j})$ if $j\in B_{\eta}(i)$ and $x_{i}=x_{j}$.
\end{conjecture}

Now, based on Conjecture \ref{conjecture}, we illustrate how GraSP and GraHTP fail to accurately identify the support of $\mathbf{x}^{*}$ from \eqref{support_identification} when $\mathbf{A}$ is highly coherent. To proceed, we consider the following example where $\mathbf{x}^{*}$ and $\hat{\mathbf{y}}$ are realized with the current estimate $\hat{\mathbf{x}}$ so as to satisfy
\begin{enumerate}
\item $i=\underset{k\in\{1, \dots, B\}}{\mathrm{argmax}}\ |\nabla h(\hat{x}_{k})|$
\item $\mathcal{J}_{\eta}(i)\cap\mathrm{supp}(\mathbf{x}^{*})=\emptyset$
\end{enumerate}
where
\begin{equation}
\mathcal{J}_{\eta}(i)=\{j\mid j\in B_{\eta}(i), \hat{x}_{i}=\hat{x}_{j}\}\setminus\{i\}
\end{equation}
is defined as the by-product of $i$. In this example, $\mathcal{J}_{\eta}(i)$ is called the by-product of $i$ because for all $j\in\mathcal{J}_{\eta}(i)$,
\begin{align}\label{example}
|\nabla h(\hat{x}_{j})|&\overset{(a)}{\approx}|\nabla h(\hat{x}_{i})|\notag\\
                       &\overset{(b)}{=}\underset{k\in\{1, \dots, B\}}{\mathrm{max}}\ |\nabla h(\hat{x}_{k})|
\end{align}
holds where (a) follows from Conjecture 1 and (b) from 1) even if $\mathcal{J}_{\eta}(i)\cap\mathrm{supp}(\mathbf{x}^{*})=\emptyset$ according to 2). The implication of this example is that most indices of $\mathcal{J}_{\eta}(i)$ are likely to be selected when $\nabla h(\hat{\mathbf{x}})$ is hard thresholded as in \eqref{support_identification} due to \eqref{example}, which results in an erroneous estimate of $\mathrm{supp}(\mathbf{x}^{*})$ because of 2). Based on the observation established from this example, we propose the BMS technique to remedy such problem.

The BMS technique provides a guideline of how to hard threshold $\nabla h(\hat{\mathbf{x}})$ in \eqref{support_identification} in order to exclude the by-product indices. The proposed BMS hard thresholding function $T_{\mathrm{BMS}, L}(\cdot)$ is an $L$-term hard thresholding function developed based on Conjecture \ref{conjecture}. The details of the BMS technique are presented in Algorithm \ref{bms}. In Line 3, the index of the maximum element of $\nabla h(\hat{\mathbf{x}})$ is selected among the unchecked index set as the current index. The by-product testing set of the current index is formed in Line 4. In Line 5, the current index is checked whether it is greater than the by-product testing set. In this paper, we refer to Line 5 as the band maximum criterion. If the current index is indeed the ``band maximum,'' which satisfies the band maximum criterion, this index is selected as the estimate of $\mathrm{supp}(\mathbf{x}^{*})$ in Line 6. Otherwise, the current index is excluded because it is likely to be the by-product of another index rather than the ground truth index. In Line 8, the unchecked index set is updated.

\begin{algorithm}[tp]
\caption{BMS hard thresholding technique}\label{bms}
\begin{algorithmic}[1]
\Require $\hat{\mathbf{x}}$, $\nabla h(\hat{\mathbf{x}})$, $L$
\Ensure $T_{\mathrm{BMS}, L}(\nabla h(\hat{\mathbf{x}}))$
\State $\mathcal{S}=\emptyset$, $\mathcal{I}=\{1, \dots, B\}$
\While {$|\mathcal{S}|<L$}
\State $i=\underset{j\in\mathcal{I}}{\mathrm{argmax}}\ |\nabla h(\hat{x}_{j})|$
\State $\mathcal{J}_{\eta}(i)=\{j\mid j\in B_{\eta}(i), \hat{x}_{i}=\hat{x}_{j}\}\setminus\{i\}$
\If {$|\nabla h(\hat{x}_{i})|>\underset{j\in\mathcal{J}_{\eta}(i)}{\mathrm{max}}\ |\nabla h(\hat{x}_{j})|$}
\State $\mathcal{S}=\mathcal{S}\cup\{i\}$
\EndIf
\State $\mathcal{I}=\mathcal{I}\setminus\{i\}$
\EndWhile
\State $T_{\mathrm{BMS}, L}(\nabla h(\hat{\mathbf{x}}))=\nabla h(\hat{\mathbf{x}})|_{\mathcal{S}}$
\end{algorithmic}
\end{algorithm}

At this point, we emphasize that Algorithm \ref{bms} is applied to $\nabla h(\hat{\mathbf{x}})$. To apply the BMS hard thresholding function to $\hat{\mathbf{x}}+\kappa\nabla h(\hat{\mathbf{x}})$ where $\kappa$ is the step size, simply replace $\nabla h(\hat{\mathbf{x}})$ with $\hat{\mathbf{x}}+\kappa\nabla h(\hat{\mathbf{x}})$ in the input, output, and Lines 3, 5, and 10 of Algorithm \ref{bms}. This variant can be derived based on the same logic using Conjecture \ref{conjecture}. Now, the BMSGraSP and BMSGraHTP algorithms are proposed to solve \eqref{map}.

BMSGraSP and BMSGraHTP are the variants of GraSP and GraHTP. The difference between our BMS-based and non-BMS-based algorithms is that $T_{\mathrm{BMS}, L}(\cdot)$ is used as a hard thresholder instead of the naive best $L$-term approximation as in \eqref{support_identification}. The details of the proposed BMSGraSP and BMSGraHTP are given in Algorithms \ref{bmsgrasp} and \ref{bmsgrahtp}. Lines 3, 4, and 5 of Algorithms \ref{bmsgrasp} and \ref{bmsgrahtp} proceed based on the same logic. In Line 3, the gradient of the objective function is computed. Then, $\mathcal{I}$ is selected from the support of the hard thresholded gradient of the objective function in Line 4, which corresponds to joint AoA and AoD estimation. In Line 5, the objective function is maximized subject to the support constraint, which can be interpreted as path gain estimation. This step can be solved via convex optimization since the objective function is concave with the support constraint being convex. Additionally, Line 6 of Algorithm \ref{bmsgrasp} hard thresholds $\mathbf{b}$ since the sparsity of $\mathbf{b}$ is at most $3L$. A natural halting condition for Algorithms \ref{bmsgrasp} and \ref{bmsgrahtp} is to halt when $\mathrm{supp}(\hat{\mathbf{x}})$ does not change from iteration to iteration \cite{bahmani2013greedy, yuan2017gradient}.

\textbf{Remark 1:} Instead of merely hard thresholding $\mathbf{b}$ in Line 6 of Algorithm \ref{bmsgrasp}, we can solve the following convex optimization problem
\begin{equation}\label{debiasing}
\hat{\mathbf{x}}=\underset{\mathbf{x}\in\mathbb{C}^{B}}{\mathrm{argmax}}\ h(\mathbf{x})\enspace\text{s.t.}\enspace\mathrm{supp}(\mathbf{x})\subseteq\mathrm{supp}(\mathbf{b}|_{L})
\end{equation}
to update $\hat{\mathbf{x}}$. This is called the debiasing variant of Algorithm \ref{bmsgrasp}, which produces a more accurate $\mathbf{x}^{*}$ \cite{bahmani2013greedy}. The complexity, however, increases.

\textbf{Remark 2:} The convex optimization problems in BMSGraSP and BMSGraHTP, which consist of Line 5 of Algorithms \ref{bmsgrasp} and \ref{bmsgrahtp}, can be solved with relatively low complexity because the support of their optimization variables are constrained to $\mathcal{I}$ where $|\mathcal{I}|=O(L)$ is typically much smaller than $B\geq MN$ in mmWave massive MIMO systems; $L$ is small due to the small number of paths, whereas $M$ and $N$ are large due to the large arrays. The same logic holds for \eqref{debiasing}. Therefore, the complexity of Algorithms \ref{bmsgrasp} and \ref{bmsgrahtp} is dominated by Line 3, which requires the gradient of the objective function defined on $\mathbb{C}^{B}$. We can reduce the complexity in Line 3 of Algorithms \ref{bmsgrasp} and \ref{bmsgrahtp} based on the fast Fourier transform (FFT) implementation of $\nabla h(\hat{\mathbf{x}})$ when $\mathbf{S}$ has an FFT-friendly structure, e.g., circularly shifted Zadoff-Chu (ZC) sequences or partial DFT matrix. From \eqref{gradient}, note that the matrix-vector multiplications involving $\mathbf{A}$ and $\mathbf{A}^{\mathrm{H}}$ act as computational bottlenecks. These matrix-vector multiplications can be efficiently implemented using the FFT by noting that
\begin{align}
             \mathrm{unvec}(\mathbf{A}\mathbf{x})&=\mathbf{A}_{\mathrm{RX}}\mathbf{X}\mathbf{A}_{\mathrm{TX}}^{\mathrm{H}}\mathbf{S}\notag\\
                                                 &=\mathbf{A}_{\mathrm{RX}}(\mathbf{S}^{\mathrm{H}}(\mathbf{A}_{\mathrm{TX}}\mathbf{X}^{\mathrm{H}}))^{\mathrm{H}},\\
\mathrm{unvec}(\mathbf{A}^{\mathrm{H}}\mathbf{c})&=\mathbf{A}_{\mathrm{RX}}^{\mathrm{H}}\mathbf{C}\mathbf{S}^{\mathrm{H}}\mathbf{A}_{\mathrm{TX}}\notag\\
                                                 &=\mathbf{A}_{\mathrm{RX}}^{\mathrm{H}}(\mathbf{A}_{\mathrm{TX}}^{\mathrm{H}}(\mathbf{S}\mathbf{C}^{\mathrm{H}}))^{\mathrm{H}}
\end{align}
where $\mathbf{C}=\mathrm{unvec}(\mathbf{c})$. By performing matrix multiplications involving $\mathbf{A}_{\mathrm{RX}}$, $\mathbf{A}_{\mathrm{TX}}$, and $\mathbf{S}$ using the FFT, the cost of computing $\nabla h(\hat{\mathbf{x}})$ can be reduced significantly.

\begin{algorithm}[tp]
\caption{BMSGraSP algorithm}\label{bmsgrasp}
\begin{algorithmic}[1]
\Require $h(\cdot)$, $L$
\Ensure $\hat{\mathbf{x}}$
\State $\hat{\mathbf{x}}=\mathbf{0}_{B}$
\While {halting condition}
\State $\mathbf{z}=\nabla h(\hat{\mathbf{x}})$
\State $\mathcal{I}=\mathrm{supp}(T_{\mathrm{BMS}, 2L}(\mathbf{z}))\cup\mathrm{supp}(\hat{\mathbf{x}})$
\State $\mathbf{b}=\underset{\mathbf{x}\in\mathbb{C}^{B}}{\mathrm{argmax}}\ h(\mathbf{x})\enspace\text{s.t.}\enspace\mathrm{supp}(\mathbf{x})\subseteq\mathcal{I}$
\State $\hat{\mathbf{x}}=\mathbf{b}|_{L}$
\EndWhile
\end{algorithmic}
\end{algorithm}

\begin{algorithm}[tp]
\caption{BMSGraHTP algorithm}\label{bmsgrahtp}
\begin{algorithmic}[1]
\Require $h(\cdot)$, $L$
\Ensure $\hat{\mathbf{x}}$
\State $\hat{\mathbf{x}}=\mathbf{0}_{B}$
\While {halting condition}
\State $\mathbf{z}=\hat{\mathbf{x}}+\kappa\nabla h(\hat{\mathbf{x}})$
\State $\mathcal{I}=\mathrm{supp}(T_{\mathrm{BMS}, L}(\mathbf{z}))$
\State $\hat{\mathbf{x}}=\underset{\mathbf{x}\in\mathbb{C}^{B}}{\mathrm{argmax}}\ h(\mathbf{x})\enspace\text{s.t.}\enspace\mathrm{supp}(\mathbf{x})\subseteq\mathcal{I}$
\EndWhile
\end{algorithmic}
\end{algorithm}

\section{Simulation Results}\label{simulation_results}
In this section, the performance of BMSGraSP and BMSGraHTP are evaluated in terms of the accuracy and achievable rate. We consider a mmWave massive MIMO system with one-bit ADCs where $M=N=64$ and $T=80$. The rows of the training signal $\mathbf{S}$ are chosen as circularly shifted ZC sequences of length $T$ \cite{1054840}, and the number of paths are $L=4$. We choose $B_{\mathrm{RX}}=256\gg M$ and $B_{\mathrm{TX}}=256\gg N$ for the proposed BMSGraSP and BMSGraHTP to reduce the leakage effect, but other channel estimators may use smaller $B_{\mathrm{RX}}$ and $B_{\mathrm{TX}}$ to prevent the breakdown caused by the resulting ill-conditioned sensing matrix. The accuracy of a channel estimator is measured based on its normalized MSE (NMSE), which is defined as
\begin{equation}
\mathrm{NMSE}=\mathbb{E}\left\{\frac{\|\hat{\mathbf{H}}-\mathbf{H}\|_{\mathrm{F}}^{2}}{\|\mathbf{H}\|_{\mathrm{F}}^{2}}\right\}
\end{equation}
where $\hat{\mathbf{H}}=\mathbf{A}_{\mathrm{RX}}\hat{\mathbf{X}}\mathbf{A}_{\mathrm{TX}}$ with $\hat{\mathbf{X}}=\mathrm{unvec}(\hat{\mathbf{x}})$.

\begin{figure}[tp]
\centering
\includegraphics[width=0.9\columnwidth]{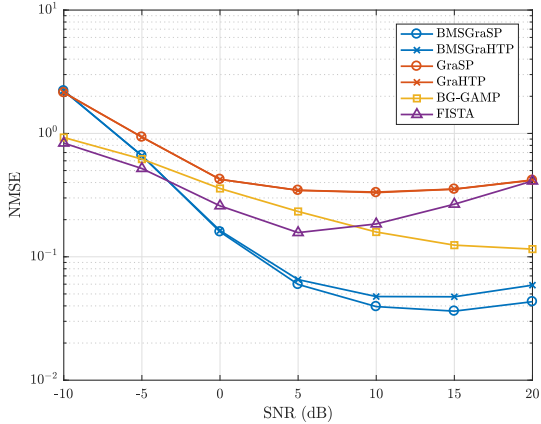}
\caption{NMSE vs. SNR with $M=N=64$, $T=80$, and $L=4$, while $B_{\mathrm{RX}}$ and $B_{\mathrm{TX}}$ vary from algorithm to algorithm.}\label{figure_1}
\end{figure}

For BMSGraSP, we consider its debiasing variant, which replaces Line 6 of Algorithm \ref{bmsgrasp} with \eqref{debiasing}. We set Algorithms \ref{bmsgrasp} and \ref{bmsgrahtp} to halt when $\mathrm{supp}(\hat{\mathbf{x}})$ does not change from iteration to iteration. We use the backtracking line search \cite{boyd2004convex} to compute $\kappa$ in Line 3 of Algorithm \ref{bmsgrahtp}. Lastly, we configure $\eta$ so as to satisfy Conjecture \ref{conjecture} by selecting the maximum $\eta$ satisfying
\begin{equation}
\underset{i\in\{1, \dots, B\}}{\mathrm{min}}\ |B_{\eta}(i)|>1.
\end{equation}

\begin{figure}[tp]
\centering
\includegraphics[width=0.9\columnwidth]{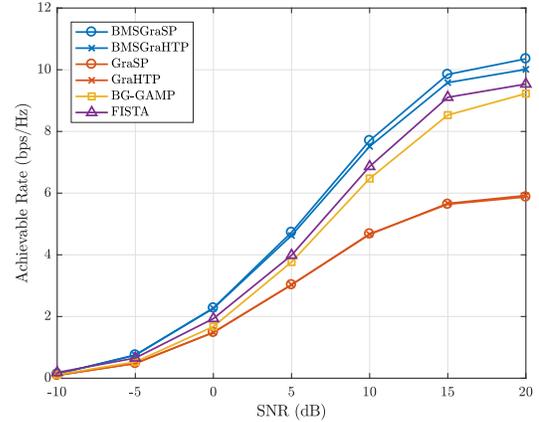}
\caption{Achievable rate lower bound \cite{8171203} vs. SNR with $M=N=64$, $T=80$, and $L=4$, while $B_{\mathrm{RX}}$ and $B_{\mathrm{TX}}$ vary from algorithm to algorithm.}\label{figure_2}
\end{figure}

The other channel estimators to be compared are GraSP \cite{bahmani2013greedy} and GraHTP \cite{yuan2017gradient}, which are non-BMS-based gradient pursuit-based algorithms, the Bernoulli-Gaussian-GAMP (BG-GAMP) \cite{8171203}, and fast iterative shrinkage-thresholding algorithm (FISTA) \cite{beck2009fast}. BG-GAMP is a loopy BP-based iterative approximate MMSE channel estimator, which assumes that $\mathbf{x}^{*}$ is distributed as i.i.d. BG, meaning that each elements are $\mathcal{CN}(0, 1)$ with probability $L/B$ but zero otherwise. FISTA is an accelerated proximal gradient descent method-based iterative MAP channel estimator based on the assumption that the logarithm of the PDF of $\mathbf{x}^{*}$ is $g_{\mathrm{FISTA}}(\mathbf{x})=-\gamma\|\mathbf{x}\|_{1}$ without the constant factor, which is the Laplace distribution. Then, the FISTA estimate of $\mathbf{x}^{*}$ is
\begin{equation}
\underset{\mathbf{x}\in\mathbb{C}^{B}}{\mathrm{argmax}}\ (f(\mathbf{x})+g_{\mathrm{FISTA}}(\mathbf{x})),
\end{equation}
which can be interpreted as the generalized LASSO \cite{tibshirani1996regression} with arbitrary forms of objective functions. For a fair comparison, we configure the regularization parameter $\gamma$ so that the expected sparsity of the FISTA estimate is $3L$, a criterion suggested in \cite{bahmani2013greedy}. Since GraSP, GraHTP, and BG-GAMP break down when the sensing matrix is ill-conditioned, we set $B_{\mathrm{RX}}=B_{\mathrm{TX}}=64$ for these channel estimators. For FISTA, we choose $B_{\mathrm{RX}}=B_{\mathrm{TX}}=256$ as in our BMSGraSP and BMSGraHTP.

The NMSE of various channel estimators is shown in Fig. \ref{figure_1} at different SNRs. In Fig. \ref{figure_1}, our channel estimators outperform other channel estimators in the medium and high SNR regimes. The poor performance of GraSP, GraHTP, and BG-GAMP is caused by the leakage effect due to the small $B_{\mathrm{RX}}$ and $B_{\mathrm{TX}}$, but increasing these parameters is forbidden since these channel estimators diverge when $\mathbf{A}$ is highly coherent. FISTA performs poorly because the Laplace distribution deviates from the true distribution of $\mathbf{x}^{*}$. In contrast, since $B_{\mathrm{RX}}$ and $B_{\mathrm{TX}}$ are large, BMSGraSP and BMSGraHTP do not suffer from the leakage effect. As a side note, we mention that all channel estimators experience performance degradation as the SNR enters the high SNR regime. This phenomenon is due to the coarse quantization of one-bit ADCs, which results in magnitude information loss. To illustrate this phenomenon, note that $\mathbf{x}^{*}$ and $c\mathbf{x}^{*}$ are indistinguishable in the high SNR regime for $c>0$ because
\begin{align}
\mathrm{Q}(\sqrt{\rho}\mathbf{A}\mathbf{x}^{*}+\mathbf{n})&\overset{(a)}{\approx}\mathrm{Q}(\sqrt{\rho}\mathbf{A}\mathbf{x}^{*})\notag\\
                                                          &\overset{(b)}{=}\mathrm{Q}(c\sqrt{\rho}\mathbf{A}\mathbf{x}^{*})
\end{align}
where (a) and (b) follow from the high SNR regime assumption and \eqref{quantization}, which implies that the information in $c$ is lost. To combat such performance degradation, the concept of dithering was suggested in \cite{7437384}, but this is beyond the scope of this paper.

We also compare the achievable rate lower bound of various channel estimators in Fig. \ref{figure_2} at different SNRs. This achievable rate lower bound, which was derived in \cite{8171203}, is obtained by selecting the precoders and decoders based on $\hat{\mathbf{H}}$ and applying the Bussgang decomposition \cite{bussgang1952crosscorrelation} in conjunction with the fact that the Gaussian noise is the worst-case noise. According to Fig. \ref{figure_2}, our channel estimators outperform other channel estimators, which agrees with Fig. \ref{figure_1}.

\section{Conclusion}
In this paper, we proposed gradient pursuit-based iterative approximate MAP channel estimators for mmWave massive MIMO systems with one-bit ADCs. In the mmWave band, the MAP channel estimation framework can be cast into a sparsity-constrained optimization problem, which is NP-hard to solve. To approximately solve such problem iteratively by pursuing the gradient of the objective function, the GraSP and GraHTP algorithms were proposed in the field of CS, which generalize the well-known CoSaMP and HTP algorithms. GraSP and GraHTP, however, break down when the objective function is ill-conditioned, which is likely to occur in the mmWave band. As a solution to such breakdown, the BMS technique, which hard thresholds the gradient of the objective function based on the band maximum criterion, was proposed in this paper. The BMS technique was applied to GraSP and GraHTP to produce the proposed BMSGraSP and BMSGraHTP algorithms. The simulation results showed that the BMSGraSP and BMSGraHTP algorithms outperform other channel estimators.

\section*{Acknowledgment}
This work was partly supported by Institute for Information \& communications Technology Promotion(IITP) grant funded by the Korea government(MSIT) (No. 2016-0-00123, Development of Integer-Forcing MIMO Transceivers for 5G \& Beyond Mobile Communication Systems) and by the National Research Foundation(NRF) grant funded by the MSIT of the Korea government (No. 2018R1A4A1025679).

\bibliographystyle{IEEEtran}
\bibliography{refs_all}

\begin{thebibliography}{10}
\providecommand{\url}[1]{#1}
\csname url@samestyle\endcsname
\providecommand{\newblock}{\relax}
\providecommand{\bibinfo}[2]{#2}
\providecommand{\BIBentrySTDinterwordspacing}{\spaceskip=0pt\relax}
\providecommand{\BIBentryALTinterwordstretchfactor}{4}
\providecommand{\BIBentryALTinterwordspacing}{\spaceskip=\fontdimen2\font plus
\BIBentryALTinterwordstretchfactor\fontdimen3\font minus
  \fontdimen4\font\relax}
\providecommand{\BIBforeignlanguage}[2]{{%
\expandafter\ifx\csname l@#1\endcsname\relax
\typeout{** WARNING: IEEEtran.bst: No hyphenation pattern has been}%
\typeout{** loaded for the language `#1'. Using the pattern for}%
\typeout{** the default language instead.}%
\else
\language=\csname l@#1\endcsname
\fi
#2}}
\providecommand{\BIBdecl}{\relax}
\BIBdecl

\bibitem{6894453}
A.~L. Swindlehurst, E.~Ayanoglu, P.~Heydari, and F.~Capolino,
  ``{Millimeter-wave massive MIMO: the next wireless revolution?}'' \emph{IEEE
  Communications Magazine}, vol.~52, no.~9, pp. 56--62, Sept. 2014.

\bibitem{6515173}
T.~S. Rappaport, S.~Sun, R.~Mayzus, H.~Zhao, Y.~Azar, K.~Wang, G.~N. Wong,
  J.~K. Schulz, M.~Samimi, and F.~Gutierrez, ``{Millimeter wave mobile
  communications for 5G cellular: it will work!}'' \emph{IEEE Access}, vol.~1,
  pp. 335--349, 2013.

\bibitem{6736746}
F.~Boccardi, R.~W. Heath, A.~Lozano, T.~L. Marzetta, and P.~Popovski, ``{Five
  disruptive technology directions for 5G},'' \emph{IEEE Communications
  Magazine}, vol.~52, no.~2, pp. 74--80, Feb. 2014.

\bibitem{6732923}
S.~Rangan, T.~S. Rappaport, and E.~Erkip, ``{Millimeter-wave cellular wireless
  networks: potentials and challenges},'' \emph{Proceedings of the IEEE}, vol.
  102, no.~3, pp. 366--385, Mar. 2014.

\bibitem{1550190}
B.~Le, T.~W. Rondeau, J.~H. Reed, and C.~W. Bostian, ``{Analog-to-digital
  converters},'' \emph{IEEE Signal Processing Magazine}, vol.~22, no.~6, pp.
  69--77, Nov. 2005.

\bibitem{7600443}
C.~Moll\'en, J.~Choi, E.~G. Larsson, and R.~W. Heath, ``{Uplink performance of
  wideband massive MIMO with one-bit ADCs},'' \emph{IEEE Transactions on
  Wireless Communications}, vol.~16, no.~1, pp. 87--100, Jan. 2017.

\bibitem{7307134}
L.~Fan, S.~Jin, C.~Wen, and H.~Zhang, ``{Uplink achievable rate for massive
  MIMO systems with low-resolution ADC},'' \emph{IEEE Communications Letters},
  vol.~19, no.~12, pp. 2186--2189, Dec. 2015.

\bibitem{7420605}
J.~Zhang, L.~Dai, S.~Sun, and Z.~Wang, ``{On the spectral efficiency of massive
  MIMO systems with low-resolution ADCs},'' \emph{IEEE Communications Letters},
  vol.~20, no.~5, pp. 842--845, May 2016.

\bibitem{7894211}
S.~Jacobsson, G.~Durisi, M.~Coldrey, U.~Gustavsson, and C.~Studer,
  ``{Throughput analysis of massive MIMO uplink with low-resolution ADCs},''
  \emph{IEEE Transactions on Wireless Communications}, vol.~16, no.~6, pp.
  4038--4051, Jun. 2017.

\bibitem{8310593}
Y.~Ding, S.~Chiu, and B.~D. Rao, ``{Bayesian channel estimation algorithms for
  massive MIMO systems with hybrid analog-digital processing and low-resolution
  ADCs},'' \emph{IEEE Journal of Selected Topics in Signal Processing},
  vol.~12, no.~3, pp. 499--513, Jun. 2018.

\bibitem{8320852}
H.~He, C.~Wen, and S.~Jin, ``{Bayesian optimal data detector for hybrid mmWave
  MIMO-OFDM systems with low-resolution ADCs},'' \emph{IEEE Journal of Selected
  Topics in Signal Processing}, vol.~12, no.~3, pp. 469--483, Jun. 2018.

\bibitem{8171203}
J.~Mo, P.~Schniter, and R.~W. Heath, ``{Channel estimation in broadband
  millimeter wave MIMO systems with few-bit ADCs},'' \emph{IEEE Transactions on
  Signal Processing}, vol.~66, no.~5, pp. 1141--1154, Mar. 2018.

\bibitem{bahmani2013greedy}
S.~Bahmani, B.~Raj, and P.~T. Boufounos, ``{Greedy sparsity-constrained
  optimization},'' \emph{Journal of Machine Learning Research}, vol.~14, no.
  Mar, pp. 807--841, 2013.

\bibitem{yuan2017gradient}
X.-T. Yuan, P.~Li, and T.~Zhang, ``{Gradient hard thresholding pursuit.}''
  \emph{Journal of Machine Learning Research}, vol.~18, pp. 166--1, 2017.

\bibitem{6965800}
T.~A. Thomas, H.~C. Nguyen, G.~R. MacCartney, and T.~S. Rappaport, ``{3D mmWave
  channel model proposal},'' in \emph{2014 IEEE 80th Vehicular Technology
  Conference (VTC2014-Fall)}, Sept. 2014, pp. 1--6.

\bibitem{1033686}
A.~M. Sayeed, ``{Deconstructing multiantenna fading channels},'' \emph{IEEE
  Transactions on Signal Processing}, vol.~50, no.~10, pp. 2563--2579, Oct.
  2002.

\bibitem{7439790}
J.~Choi, J.~Mo, and R.~W. Heath, ``{Near maximum-likelihood detector and
  channel estimator for uplink multiuser massive MIMO systems with one-bit
  ADCs},'' \emph{IEEE Transactions on Communications}, vol.~64, no.~5, pp.
  2005--2018, May 2016.

\bibitem{7094595}
J.~Mo, P.~Schniter, N.~G. Prelcic, and R.~W. Heath, ``{Channel estimation in
  millimeter wave MIMO systems with one-bit quantization},'' in \emph{2014 48th
  Asilomar Conference on Signals, Systems and Computers}, Nov. 2014, pp.
  957--961.

\bibitem{needell2009cosamp}
D.~Needell and J.~A. Tropp, ``{CoSaMP: iterative signal recovery from
  incomplete and inaccurate samples},'' \emph{Applied and computational
  harmonic analysis}, vol.~26, no.~3, pp. 301--321, 2009.

\bibitem{foucart2011hard}
S.~Foucart, ``{Hard thresholding pursuit: an algorithm for compressive
  sensing},'' \emph{SIAM Journal on Numerical Analysis}, vol.~49, no.~6, pp.
  2543--2563, 2011.

\bibitem{boyd2004convex}
S.~Boyd and L.~Vandenberghe, \emph{{Convex optimization}}.\hskip 1em plus 0.5em
  minus 0.4em\relax Cambridge university press, 2004.

\bibitem{fannjiang2012coherence}
A.~Fannjiang and W.~Liao, ``{Coherence pattern--guided compressive sensing with
  unresolved grids},'' \emph{SIAM Journal on Imaging Sciences}, vol.~5, no.~1,
  pp. 179--202, 2012.

\bibitem{1577905}
N.~Jindal, ``{MIMO broadcast channels with finite rate feedback},'' in
  \emph{GLOBECOM '05. IEEE Global Telecommunications Conference, 2005.},
  vol.~3, Nov. 2005, pp. 5 pp.--.

\bibitem{shrivastava2014asymmetric}
A.~Shrivastava and P.~Li, ``{Asymmetric LSH (ALSH) for sublinear time maximum
  inner product search (MIPS)},'' in \emph{Advances in Neural Information
  Processing Systems}, 2014, pp. 2321--2329.

\bibitem{1054840}
D.~Chu, ``{Polyphase codes with good periodic correlation properties
  (Corresp.)},'' \emph{IEEE Transactions on Information Theory}, vol.~18,
  no.~4, pp. 531--532, Jul. 1972.

\bibitem{beck2009fast}
A.~Beck and M.~Teboulle, ``{A fast iterative shrinkage-thresholding algorithm
  for linear inverse problems},'' \emph{SIAM journal on imaging sciences},
  vol.~2, no.~1, pp. 183--202, 2009.

\bibitem{tibshirani1996regression}
R.~Tibshirani, ``{Regression shrinkage and selection via the lasso},''
  \emph{Journal of the Royal Statistical Society. Series B (Methodological)},
  pp. 267--288, 1996.

\bibitem{7437384}
N.~Liang and W.~Zhang, ``{Mixed-ADC massive MIMO},'' \emph{IEEE Journal on
  Selected Areas in Communications}, vol.~34, no.~4, pp. 983--997, Apr. 2016.

\bibitem{bussgang1952crosscorrelation}
J.~J. Bussgang, ``{Crosscorrelation functions of amplitude-distorted Gaussian
  signals},'' 1952.

\end{thebibliography}

\end{document}